\documentclass[twocolumn,showpacs,amsfonts,aps,prl,floatfix,%
groupedaddress]{revtex4}

\usepackage{amsmath}
\usepackage{graphicx}
\usepackage{bm}

\newcommand{\vect}[1]{\bm{#1}}

\newcommand{\cf}{\emph{cf.}}
\newcommand{\ptc}{p_\mathrm{cut}}

\def\eq#1{{Eq.~(\ref{#1})}}
\def\fig#1{{Fig.~\ref{#1}}}
\newcommand{\be}{\begin{equation}}
\newcommand{\ee}{\end{equation}}

\graphicspath{{Figs/}}

\begin{document}

\title{Gluon Saturation Effects in Relativistic U+U Collisions}
\author{Anthony Kuhlman}
\affiliation{Department of Physics, The Ohio State University, 
  Columbus, OH 43210, USA}
\author{Ulrich Heinz}
\email[Correspond to\ ]{heinz@mps.ohio-state.edu}
\affiliation{Department of Physics, The Ohio State University, 
  Columbus, OH 43210, USA}
\author{Yuri V. Kovchegov}
\affiliation{Department of Physics, The Ohio State University, 
  Columbus, OH 43210, USA}

\begin{abstract}
We examine entropy production in relativistic U+U collisions on the 
basis of a Color Glass Condensate (CGC) type picture as implemented
in the Kharzeev-Levin-Nardi model (KLN).  In this framework, we find 
that the peak entropy density produced in tip-on-tip U+U collisions 
is about 30\% greater than that seen in central Au+Au collisions.
Although the resulting difference in the produced charged particle 
multiplicity between tip-on-tip and side-on-side collisions is
smaller than that predicted by previous Glauber model estimates,
it is still large enough to allow for experimental discrimination
between average orientations of the uranium nuclei. We also point 
out that in the saturation/CGC approach the collision geometry plays 
a more important role than previously believed, and that the observed
centrality dependence of the produced particle multiplicity per 
participant in Au+Au collisions can be qualitatively reproduced 
even without running coupling effects.
\end{abstract}

\date{\today}


\maketitle

Recent studies \cite{Heinz:2004ir,Kuhlman:2005ts,Nepali:2006va} have 
pushed for a uranium-uranium (U+U) collision program at RHIC 
to address a number of unanswered questions about the quark-gluon plasma 
(QGP). In particular, it is thought \cite{Heinz:2004ir,Kuhlman:2005ts}
that the higher entropy densities produced in U+U over those found in 
Au+Au collisions would provide a definitive test of the seemingly ideal 
hydrodynamic behavior of the elliptic flow observed at RHIC
\cite{Whitepapers,QGP3v2}.

Previous estimates \cite{Heinz:2004ir,Kuhlman:2005ts,Nepali:2006va} 
relied on a Glauber model calculation of the entropy densities produced 
in U+U collisions. These results showed a substantial increase in the 
initially produced entropy density over that seen in even the
most central Au+Au collisions. Perhaps most importantly, a clear
($\sim 15\%$) difference in the multiplicities produced in tip-on-tip 
versus side-on-side U+U collisions allowed the selection of average 
relative orientations of the two uranium nuclei, by placing cuts on 
the charged particle multiplicity \cite{Heinz:2004ir}. This remains
true even if experimental inefficiencies in triggering on full-overlap
U+U collisions, using strict spectator nucleon cuts, are taken into
account \cite{Kuhlman:2005ts}. 

However, our theoretical understanding of the initial particle production 
processes during the nuclear collision and the conditions of the produced
plasma at early times remains limited, and it has recently been shown
\cite{Adil:2005bb,Hirano:2005} that this results in significant ambiguities 
for the initial source eccentricities in non-central Au+Au collisions at 
RHIC. Hirano and collaborators \cite{Adil:2005bb,Hirano:2005} found that,
if one uses parametrizations for the initial entropy production that are
based on the Color Glass Condensate model \cite{Iancu:2002xk} of gluon 
saturation in large nuclei at high energies \cite{Gribov:1984tu},
in particular the Kharzeev-Levin-Nardi (KLN) parametrization
\cite{Kharzeev:2000ph,KLNmodel,Hirano:2004rs}, one obtains initial
density profiles in the transverse plane which are flatter and have
steeper edges than the more Gaussian-looking profiles \cite{Kolb:2001qz}
one obtains from the Glauber model. As a consequence, one obtains initial 
source eccentricities which, over most of the impact parameter range, are 
significantly larger in the KLN model than in the Glauber model. As shown 
in Ref.~\cite{Hirano:2005}, this ambiguity in the initial conditions has
considerable impact on the interpretation of available elliptic flow data
and the question to what extent the QGP created at RHIC behaves like a
``perfect fluid''.

In the present paper we investigate to what extent these model ambiguities
for the initial entropy production also affect the usefulness of a U+U 
collision program where one tries to exploit the intrinsic deformation 
of the colliding nuclei to generate deformed fireballs even in central
collisions. Specifically, we study the influence of various model
assumptions about the initial entropy production on the distribution
of source eccentricities in U+U collisions, in order to assess how 
model ambiguities affect our ability to select for specific collision 
geometries by cutting event samples on the number of spectator nucleons 
and on charged particle multiplicity 
\cite{Heinz:2004ir,Kuhlman:2005ts,Nepali:2006va}. 

We begin by outlining our procedure for calculating the initial entropy
production from the Color Glass Condensate model. We use the KLN 
parametrization \cite{Kharzeev:2000ph,KLNmodel}, in the specific form 
implemented by Hirano and Nara in Ref.~\cite{Hirano:2004rs}, where the 
initial energy per unit rapidity and per unit transverse area is given 
by the $k_T$-factorization formula \cite{Gribov:1984tu}
\begin{eqnarray}
\label{kt_factorization}
&&\frac{dE}{d^2x_\perp dy} = \frac{2\pi^3N_c}{N_c^2 - 1} 
\int^{p_\mathrm{cut}} \frac{d^2p_T}{p_T} \int_0^{p_T} d^2k_T
\nonumber\\
&&\qquad\qquad\quad\times\,
  \alpha\left(\max\Bigl\{\frac{(\vect{k}_T{-}\vect{p}_T)^2}{4},
                        \frac{(\vect{k}_T{+}\vect{p}_T)^2}{4}\Bigr\}\right)
\\
&&\times\,
  \phi_A\!\left(x_1,\frac{(\vect{k}_T{+}\vect{p}_T)^2}{4};\vect{x}_\perp\right)
  \phi_B\!\left(x_2,\frac{(\vect{k}_T{-}\vect{p}_T)^2}{4};\vect{x}_\perp\right)
  \,,
\nonumber
\end{eqnarray}
with $x_{1,2} = p_T\ \exp(\pm y)/\sqrt{s}$. The unintegrated gluon 
distribution for nucleus $A$ is taken to be
\begin{equation}
\label{gluon_distribution}
\phi_A\left(x,k^2;\vect{x}_\perp\right) = 
\frac{\kappa~ C_F~ Q_s^2}{2\pi^3\alpha_s(Q_s^2)}
\left\{
\begin{array}{cc}
\frac{1}{Q_s^2 + \Lambda^2}, & k \le Q_s,\\
\frac{1}{k^2 + \Lambda^2}, & k > Q_s,
\end{array}
\right.
\end{equation}
where the saturation momentum $Q_s$ is determined by solving
\begin{equation}
\label{Qs_equation}
Q_s^2(x,\vect{x}_\perp) = \frac{2\pi^2}{C_F} \alpha_s(Q_s^2)\,
xG(x,Q_s^2)\, n_{\mathrm{part},A}(\vect{x}_\perp),
\end{equation}
and similarly for nucleus $B$. The normalization $\kappa$ in 
Eq.~\eqref{gluon_distribution} is fixed by the measured charged 
particle multiplicity $dN_\mathrm{ch}/dy$ in central Au+Au collisions 
at $\sqrt{s}=200\,A$\,GeV (see below). To reproduce the results 
of Refs.~\cite{Hirano:2005} and \cite{Hirano:2004rs}, we take 
$\kappa^2 = 1.8$.

Both the energy density \eqref{kt_factorization} and the local (i.e.
$\vect{x}_\perp$-de\-pendent) saturation momenta of the two nuclei, 
Eq.~\eqref{Qs_equation}, depend on the impact parameter of the 
collision through the density of wounded nucleons $n_\mathrm{part}$ 
in the transverse plane, computed from the standard Glauber model 
formula (see, e.g., Ref.~\cite{Kolb:2001qz}). For nucleus $A$ it is 
given by 
\begin{eqnarray}
\label{npart_equation}
&& n_{\mathrm{part},A}(\vect{r}_\perp,\vect{b}) 
   = 
\nonumber\\
&& T_A\left(\vect{r}_\perp{+}\frac{\vect{b}}{2}\right)
   \left(1 - \left(1 - \frac{\sigma_{NN}T_B
   \left(\vect{r}_\perp{-}\frac{\vect{b}}{2}\right)}{B}\right)^B\right),
   \quad
\end{eqnarray}
while for nucleus $B$ the nuclear thickness functions $T_A$ and $T_B$
should be interchanged. The latter are computed from a Woods-Saxon 
distribution for the density of the uranium nucleus, with radius 
$R(\theta) = (6.8\ \mathrm{fm})(0.91+0.26 \cos^2\theta)$ (where $\theta$
is the polar angle with the nuclear symmetry axis) and surface thickness 
$\xi=0.54$\,fm. 

The integrated gluon distribution, $xG(x,k^2)$, in Eq.~\eqref{Qs_equation} 
is given by
\begin{equation}
\label{integrated_gluon}
xG(x,k^2) = K
\ln\left(\frac{k^2+\Lambda^2}{\Lambda_{QCD}^2}\right)x^{-\lambda}(1-x)^4,
\end{equation}
with $K{\,=\,}0.7$ and $\lambda{\,=\,}0.2$ \cite{Hirano:2004rs}. 
The running coupling constant $\alpha_s$ is computed from the 
one-loop expression
\begin{equation}
\label{alphas}
\alpha_s(k^2) = \frac{4\pi}{\beta_0
  \ln\left(\frac{k^2+\Lambda^2}{\Lambda_{QCD}^2}\right)}
\end{equation}
with $\beta_0{\,=\,}11{-}\frac{2}{3} n_f{\,=\,}9$. To render the 
calculations infrared safe, we limit the growth of the running 
coupling constant at small $k^2$ by cutting it off at 
$\alpha_s(k^2){\,=\,}0.5$ \cite{Hirano:2005}. In all expressions 
above we use $\Lambda{\,=\,}\Lambda_{QCD}{\,=\,}0.2$\,GeV 
\cite{Hirano:2004rs,Hirano:2005}.

\begin{figure}[tb]
\includegraphics[width=0.45\textwidth,clip]{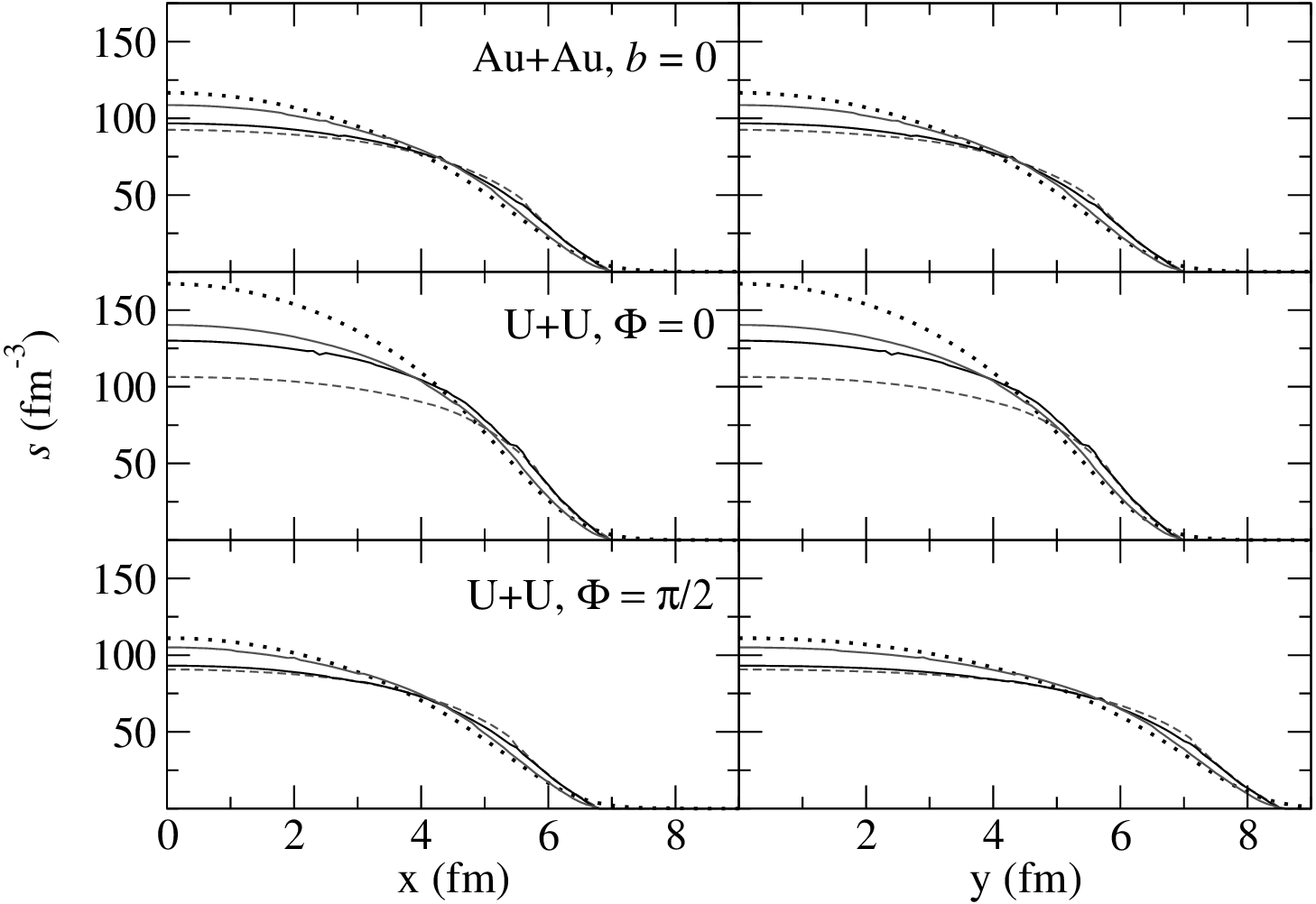}
\caption{Transverse entropy density profiles in the in-plane direction
  (left) compared with those in the out-of-plane direction (right)
  for central Au+Au (top panels), tip-on-tip U+U (center panels) and
  side-on-side U+U collisions (bottom panels). Shown are results from 
  the Glauber model (dotted) and from the KLN model with three different 
  $p_T$ cutoffs (see text): 
  $p_\mathrm{cut}{\,=\,}3$\,GeV/$c$ (dashed gray), 
  $p_\mathrm{cut}{\,=\,}2\,Q_s^\mathrm{max}$ (solid black),
  and $p_\mathrm{cut}{\,=\,}\infty$ (solid gray).} 
\label{F1}
\end{figure}

From equation \eqref{kt_factorization} we obtain the energy density at 
thermalization time $\tau_0$ (for which we use $\tau_0{\,=\,}0.6$\,fm/$c$) 
as $e(\vect{x}_\perp,\tau_0)=dE/(\tau_0\ d^2x_\perp\ d\eta)$
at $\eta{\,=\,}y{\,=\,}0$. We then translate this energy density $e$ 
into an entropy density $s$, by assuming that it thermalizes and using 
the relationship $s{\,=\,}8.67\,e^{3/4}$ (where $e$ must be inserted in 
GeV/fm$^3$) for a thermalized ideal gas of quarks and gluons. By adjusting 
$\kappa$, this entropy density is then normalized such that for impact
parameter $b{\,=\,}0$ its integral over the transverse plane yields the 
same result as the Glauber model in reference \cite{Heinz:2004ir}
(\cf\ Figure \ref{F1}).

\begin{figure}[tb]
\includegraphics[width=0.45\textwidth,clip]{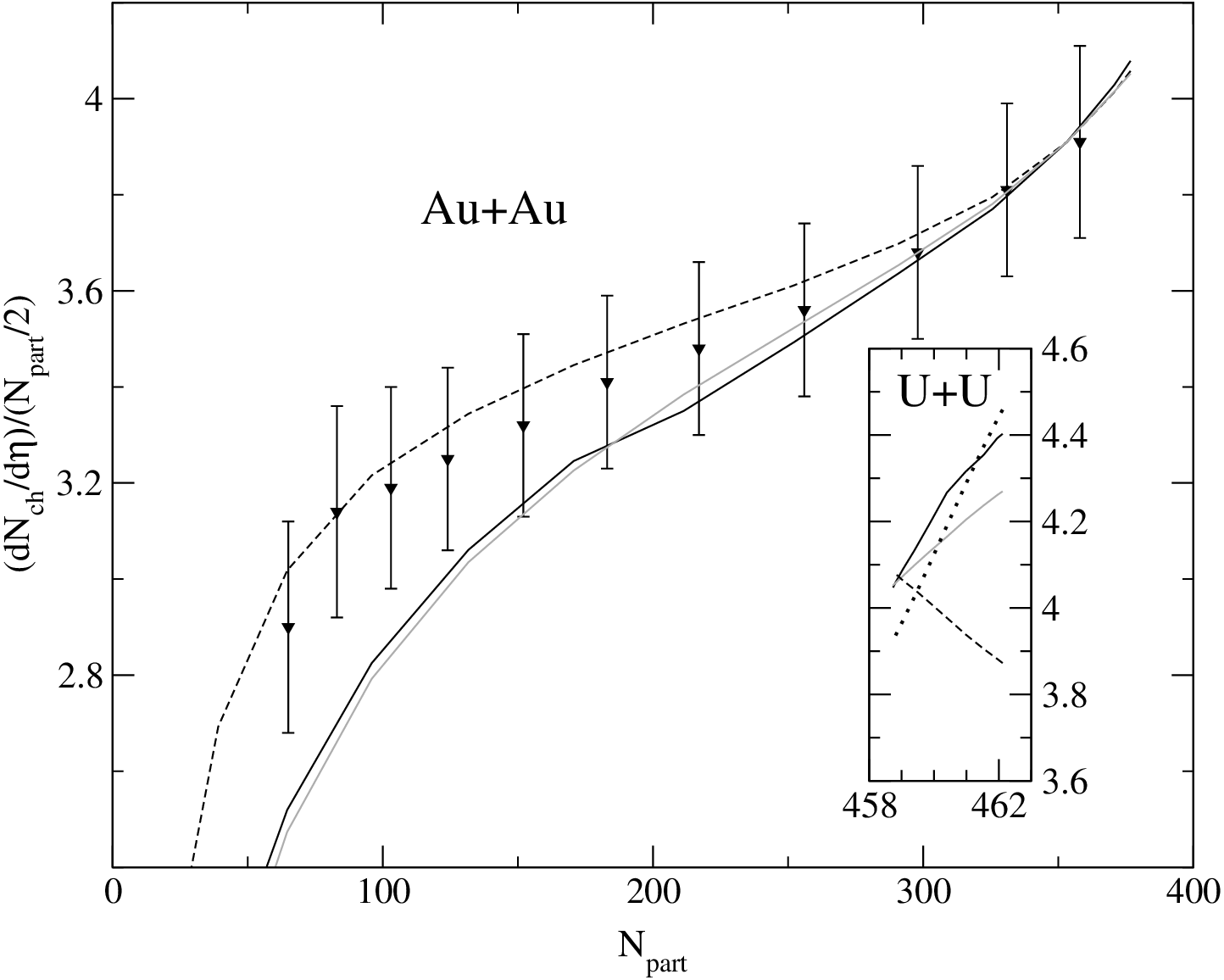}
\caption{Charged particle multiplicity per participant pair as a
  function of centrality for Au+Au. The inset shows corresponding
  predictions for full-overlap U+U collisions. The data are from 
  PHOBOS at $\sqrt{s}{\,=\,}200\,A$\.,GeV \cite{Back:2002gz}. Shown 
  are results from the Glauber model (dotted) and from the KLN model 
  different $p_T$ cutoffs: $p_\mathrm{cut}{\,=\,}3$\,GeV/$c$ 
  (dashed), $p_\mathrm{cut}{\,=\,}2\,Q_s^\mathrm{max}$ (black), 
   and $p_\mathrm{cut}{\,=\,}\infty$ (gray).}
\label{F2}
\end{figure}

We then assume that the charged particle multiplicity, $dN_{ch}/d\eta$,
is proportional to the total entropy produced in the plane,
$dN_{ch}/d\eta = \Gamma \int d^2r_\perp s(\vect{r}_\perp)$, and adjust
the proportionality constant $\Gamma=0.088$ to agree with central Au+Au 
data obtained by the PHOBOS collaboration \cite{Back:2002gz}. 
The resulting fit is shown in Figure \ref{F2}. These parameters are 
then assumed to also describe U+U collisions at the same collision 
energy.

In the work of Hirano and Nara, the integral over $p_T$ in 
Eq.~\eqref{kt_factorization} is cut off at a fixed value of 
$p_\mathrm{cut}{\,=\,}3$\,GeV, to limit the contribution from 
high-$p_T$ gluons which thermalize incompletely \cite{Hirano:2004rs}.
As shown in Fig.~\ref{F2}, the resulting centrality dependence of 
the charged particle multiplicity in 200\,$A$\,GeV Au+Au collisions 
reproduces the PHOBOS data \cite{Back:2002gz} very well. However,
the use of a fixed $p_T$ cutoff in Eq.~\eqref{kt_factorization}, 
independent of impact parameter, raises questions: The momentum 
dependence of the unintegrated gluon distributions $\phi_{A,B}$ 
is controlled by their respective saturation momenta in 
Eq.~\eqref{Qs_equation} which do depend on both $\bm{x}_\perp$
and $\bm{b}$; by implementing a fixed cutoff in $p_T$, one doesn't 
allow the $p_T$ integral to fully explore this $b$-dependence. As 
we discuss now (and show in Figure \ref{F2}), this leads to a 
significant distortion of the centrality dependence of charged hadron 
production. 

At any given point $\bm{x}_\perp$ in the overlap region of the two nuclei,
the result of equation \eqref{kt_factorization} is controlled by two
saturation momenta, $Q_{s,\mathrm{min}}$ and $Q_{s,\mathrm{max}}$, 
corresponding to the smaller and larger of the nuclear thicknesses of 
the two colliding nuclei at this point. If we ignore the running of 
$\alpha_s$, the integrand of equation \eqref{kt_factorization} is
essentially constant for $p_T{\,<\,}Q_{s,\mathrm{min}}$, decreases
like $1/p_T^2$ for $Q_{s,\mathrm{min}}{\,<\,}p_T{\,<\,}Q_{s,\mathrm{max}}$
and like $1/p_T^4$ for $p_T{\,>\,}Q_{s,\mathrm{max}}$. With a fixed
$p_T$ cutoff we cut off more of the $1/p_T^4$ tail in central than in 
more peripheral collisions, due to the fact that $Q_{s,\mathrm{max}}$ 
is larger in the former case. The net effect of this is to suppress 
entropy production in central relative to peripheral collisions. Indeed,
the centrality dependence of $dN_\mathrm{ch}/dy$ becomes noticeably 
steeper if this $p_T$ cutoff is removed (solid gray line in Figure 
\ref{F2}). The same effect can be achieved by allowing the $p_T$ cutoff 
to scale with the (centrality dependent) effective saturation momentum. 
We tested this by setting $p_\mathrm{cut}{\,=\,}2\,Q_s^\mathrm{max}(b)$ 
where we chose (somewhat arbitrarily) $Q_s^\mathrm{max}(b){\,=\,}Q_s(x{=}0.01,
\vect{x}_\perp{=}\mathbf{0};\bm{b})$ (solid black line in Figure~\ref{F2}).

Even if for Au+Au collisions the effect of removing the $p_T$ cutoff in 
Eq.~\eqref{kt_factorization} is clearly noticeable in the centrality
dependence of $dN_\mathrm{ch}/dy$, it remains quantitatively small 
($\sim15\%$ in peripheral collisions). For full-overlap U+U collisions, 
however, the different ways of cutting off or not cutting off the $p_T$ 
integral produce {\em qualitatively different} results. The inset in 
Figure \ref{F2} shows that for a fixed $p_T$ cutoff of $3$\,GeV/$c$ 
(dashed black line), tip-on-tip collisions produce {\em lower} 
multiplicities than side-on-side collisions, in stark contradiction 
with the Glauber model prediction (dotted line). The qualitative tendency 
of the Glauber model to produce higher multiplicities in tip-on-tip 
collisions \cite{Heinz:2004ir,Kuhlman:2005ts,Nepali:2006va} (due to the 
larger binary collision component in this configuration) is recovered in 
the KLN saturation model if the $p_T$ cutoff is either entirely removed
(solid gray line) or assumed to scale with the saturation momentum 
(solid black line). In either case, however, the quantitative difference
between tip-on-tip and side-on-side multiplicities is somewhat smaller 
for the KLN model than for the Glauber parametrization.

\begin{figure}[tb]
\includegraphics[width=0.45\textwidth,clip]{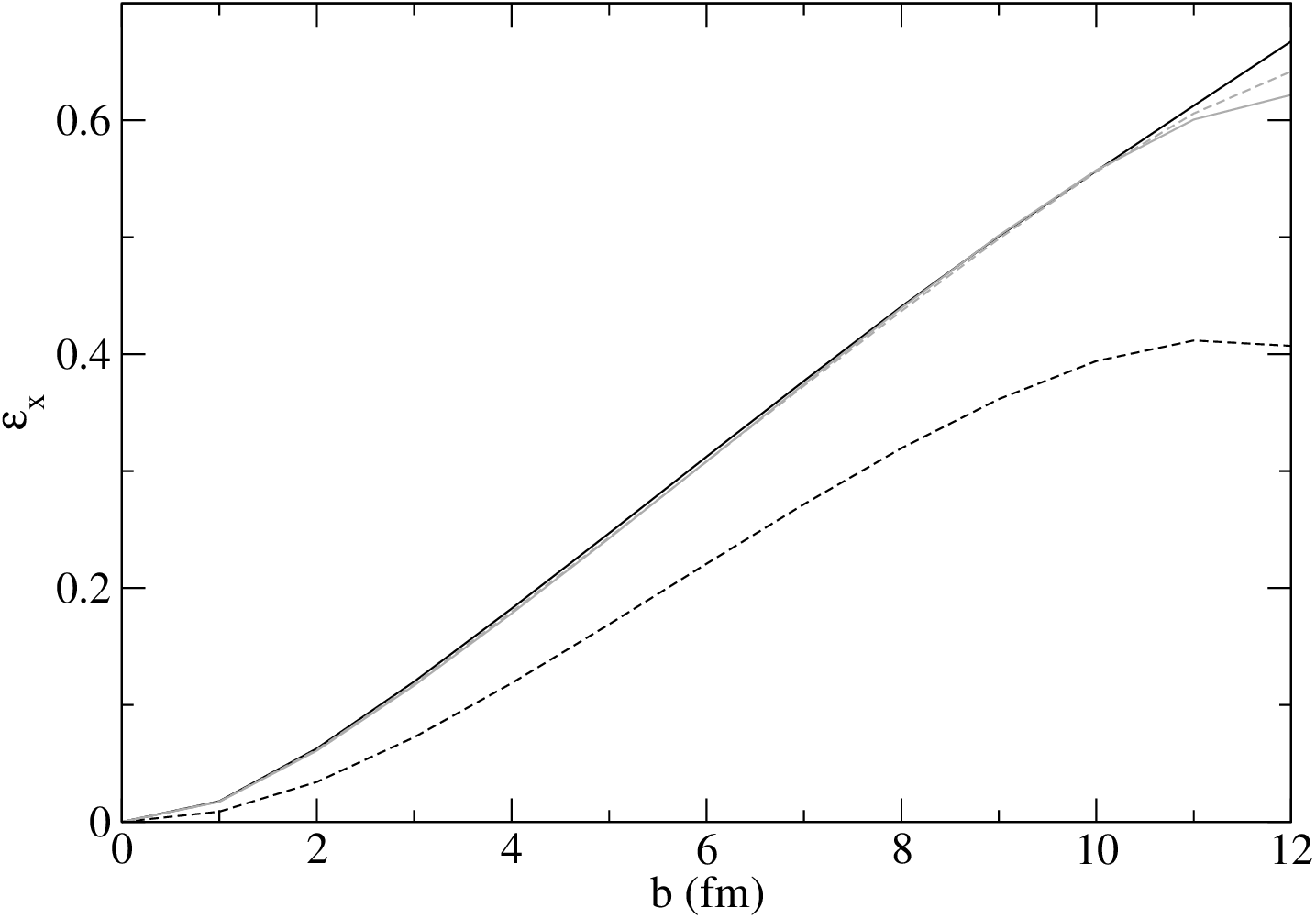}
\caption{Eccentricity versus impact parameter for Au+Au collisions at
  $\sqrt{s}=200$ GeV for the Glauber model (dashed) and the KLN model
  with $\ptc{\,=\,}2\,Q_s$ (black), $\ptc{\,=\,}\infty$ (gray) and 
  $\ptc=3$\,GeV/$c$ (dashed gray).}
\label{F3}
\end{figure}

Returning to the initial entropy density profiles shown in Figure \ref{F1},
we can now see how these differences arise: Generically, the profiles from
the saturation model are flatter near the center and steeper near the
edge than those from the Glauber model (which look more Gaussian). 
In tip-on-tip collisions (center panels) the Glauber model shows the 
strongest growth of entropy density in the middle of the overlap region, 
due to a strong increase of the binary collision component. This
growth is tempered in the KLN saturation model, due to the slower
(logarithmic) increase of the saturation momentum $Q_s$ as one approaches
the center of the overlap region. If one implements a fixed $p_T$
cutoff, the $p_T$ integral in Eq.~\eqref{kt_factorization} can not 
even fully explore this logarithmic growth of the saturation momentum, 
and the entropy density increases even more slowly towards the center. 
In fact, the difference in entropy {\em density} between tip-on-tip
and side-on-side collisions then becomes so small that it can no longer
compensate for the smaller overlap area in tip-on-tip collisions, leading
to the aforementioned counterintuitive result of {\em smaller} total 
multiplicities in tip-on-tip compared to side-on-side collisions (dashed 
gray line in the inset of Fig. \ref{F2}).

As already pointed out in Refs.~\cite{Adil:2005bb,Hirano:2005}, the 
more plateau-like entropy distributions from the KLN model yield larger 
eccentricities 
$\epsilon_x{\,=\,}\frac{\langle y^2{-}x^2\rangle}{\langle y^2{+}x^2\rangle}$
than the Glauber model profiles, due to their bigger weights at larger
distance from the fireball center. In Figure \ref{F3} we show 
the eccentricity of the initial entropy distribution as a function
of impact parameter $b$ for Au+Au collisions. We confirm the observation
made in \cite{Hirano:2005} that the KLN eccentricities exceed the Glauber
ones by about 25-30\% over most of the impact parameter range, and further
show that they are insensitive to the choice of the $p_T$ cutoff in 
Eq.~\eqref{kt_factorization}. Since the value of the initial source 
eccentricity controls the finally observed elliptic flow, Fig.~\ref{F3}
reassures us that the conclusions of Ref.~\cite{Hirano:2005} as to
the interpretation of the measured $v_2$ values are not affected by
model uncertainties related to the choice of $p_T$ cutoff.

In \cite{Heinz:2004ir,Kuhlman:2005ts} two of us advocated using
the multiplicity to select for particular orientations of the
colliding uranium nuclei. In the Glauber model framework, we found
that of all possible orientations of full-overlap U+U collisions, the
highest multiplicities were produced in tip-on-tip collisions and the
lowest multiplicities were formed in those collisions between nuclei
in the side-on-side configuration. The difference in mean multiplicity 
between these extreme cases was $\approx 14$\% \cite{Heinz:2004ir}.
This provided a ``handle'' to select for average orientations of the 
nuclei: By selecting events with high multiplicity, one effectively 
triggers on low source eccentricity, and \emph{vice versa}. The 14\% 
difference in multiplicities between tip-on-tip and side-on-side 
collisions proved large enough that such a discrimination between
collision configurations with large and small source eccentricities
remained feasible even after experimental inefficiencies in triggering
on full-overlap U+U collisions (by selecting events with a small number
of spectator nucleons) were taken into account \cite{Kuhlman:2005ts}.

The inset in Fig.~\ref{F2} shows that with CGC/KLN initial conditions
the range of multiplicities produced by the different orientations of
full-overlap collisions is somewhat smaller than in the Glauber model 
calculations. It may therefore be prudent to reanalyze the possible
degradation of our ability to select event classes of well-defined
source eccentricity resulting from trigger inefficiencies and 
event-by-event multiplicity fluctuations.

\begin{figure}[tb]
\includegraphics[width=0.45\textwidth,clip]{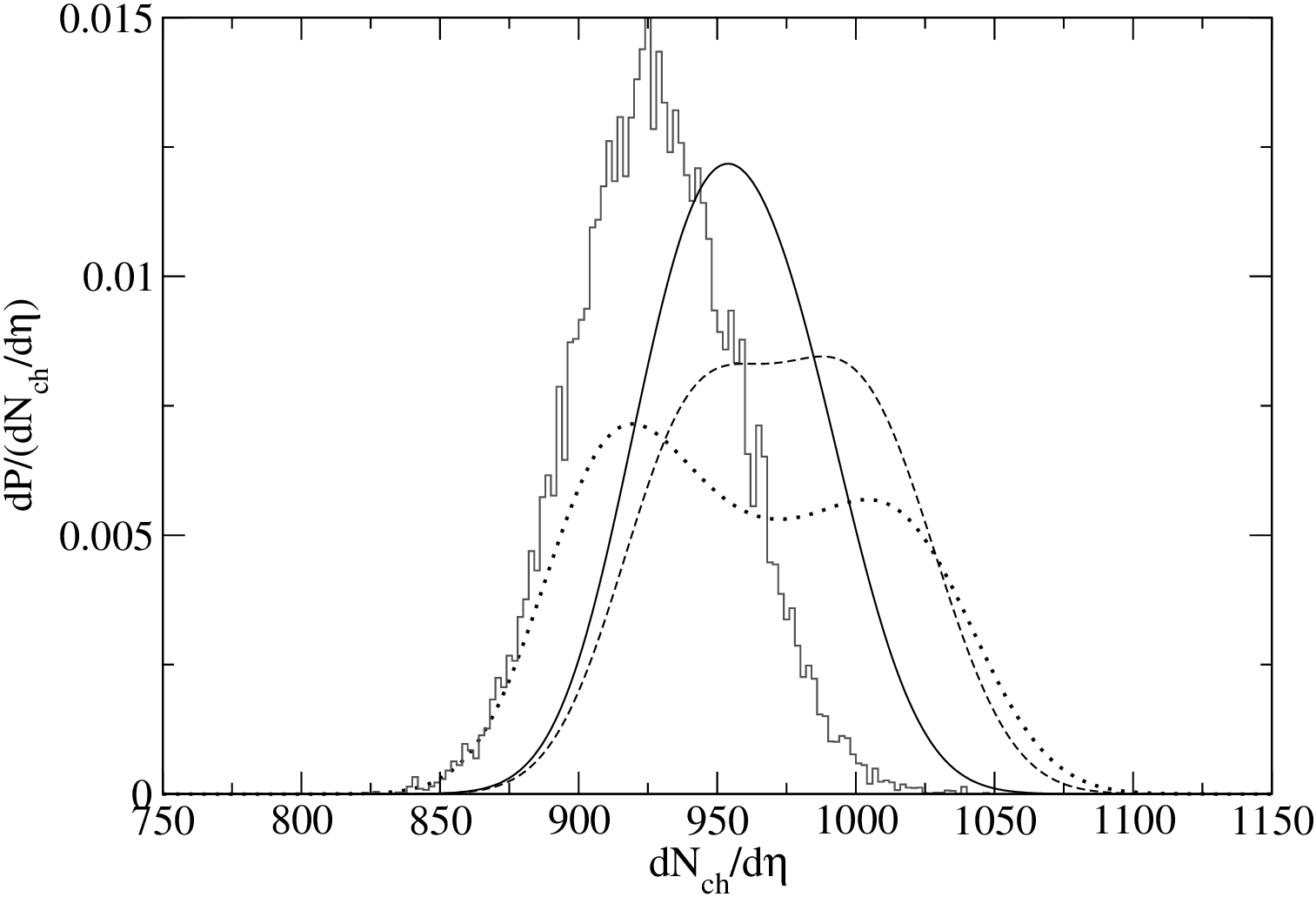}
\caption{Multiplicity distribution for full-overlap U+U collisions.
  Shown are results from the Glauber model (dotted) and the KLN model 
  with $p_\mathrm{cut}{\,=\,}2\,Q_s^\mathrm{max}$ (dashed) and 
  $p_\mathrm{cut}{\,=\,}\infty$ (solid black). Also shown is the 
  result of cutting the full multiplicity distribution given in 
  Figure~\ref{F6} on the 0.5\% of events with the lowest number of 
  spectators.}
\label{F4}
\end{figure}

We first study the effect of event-by-event multiplicity fluctuations,
assuming that full-overlap collisions can be selected with 100\% efficiency.
As before \cite{Heinz:2004ir} we adopt as probability distribution for 
the event-by-event multiplicity the expression \cite{Kharzeev:2000ph}
\begin{equation}
\label{probability_distribution}
\frac{dP}{dn\, d\Phi} = A
\exp\left\{-\frac{(n-\bar{n}(\Phi))^2}{2a\bar{n}(\Phi)}\right\},
\end{equation} 
where $n$ is the measured multiplicity in a given event and $\bar{n}(\Phi)$ 
is the average multiplicity for a given orientation $\Phi$ of the nuclei, 
now computed from the KLN model using the procedure described above. For 
the parameter $a$ controlling the width of the fluctuations we again take
$a=0.6$ as this has been shown to produce good agreement with existing
Au+Au data \cite{Kharzeev:2000ph}.

Integrating Eq.~\eqref{probability_distribution} over $\Phi$
yields the multiplicity distribution for full-overlap U+U collisions, 
shown in Fig.~\ref{F4}. As expected, the distributions obtained from 
the KLN model are considerably narrower than the one resulting from the
Glauber model. Furthermore, the shapes of the distributions are quite 
different, a result of the individual Gaussian distributions from 
tip-on-tip and side-on-side collisions being squeezed on top of one 
another, due to the smaller difference in mean multiplicity between them.

\begin{figure}[tb]
\includegraphics[width=0.45\textwidth,clip]{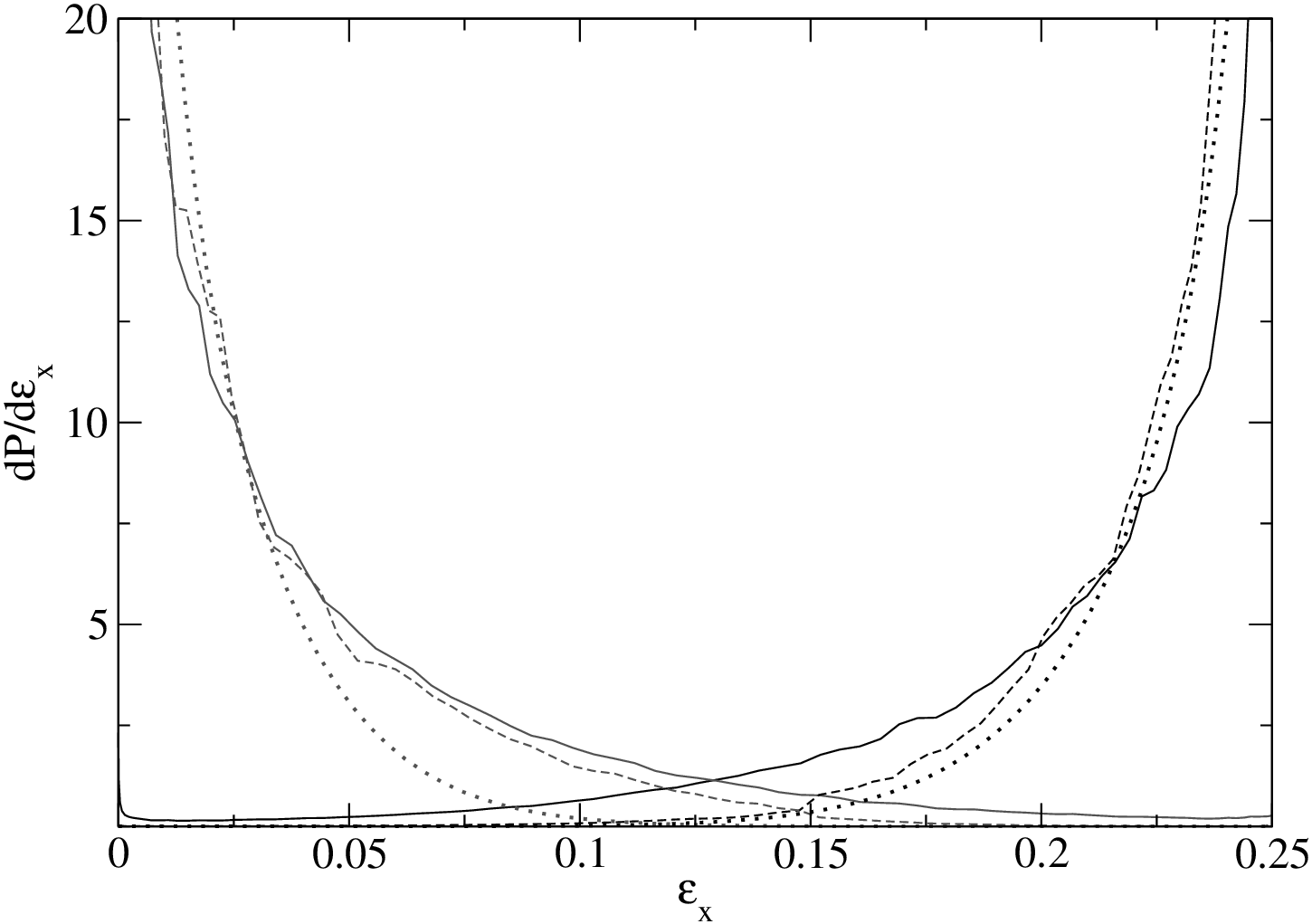}
\caption{Eccentricity distributions resulting from cutting the
  multiplicity distributions of Fig.~\ref{F4} on the 10\% of events 
  with the lowest (black) and highest (gray) multiplicities.
  Shown are results for the Glauber model (dotted) and the KLN model 
  with $p_\mathrm{cut}{\,=\,}2\,Q_s^\mathrm{max}$ (dashed) and
  $p_\mathrm{cut}{\,=\,}\infty$ (solid).}
\label{F5}
\end{figure}

After placing cuts on the multiplicity distributions of Fig.~\ref{F4}, 
we can convert the multiplicity distribution into a distribution of 
eccentricities according to
\begin{equation}
\label{probability_convert}
\left.\frac{dP}{d\epsilon_x}\right|_{n_0}^{n_1} = \int_{n_0}^{n_1} dn
\frac{dP}{dn\ d\Phi} \frac{d\Phi}{d\epsilon_x},
\end{equation}
where $n_0$ and $n_1$ represent the lower and upper limits of the
multiplicity cuts, respectively. The resulting eccentricity
distributions are shown in Fig.~\ref{F5} for cuts on events with the
10\% lowest and highest multiplicities, respectively. Not
unexpectedly, the distributions resulting from the KLN model are
slightly broader than those from the Glauber model: Due to the smaller
range of multiplicities between the $\Phi{\,=\,}0$ and
$\Phi{\,=\,}\frac{\pi}{2}$ configurations, the eccentricities
corresponding to given multiplicity slice are less sharply defined.
This also results in a slightly higher (lower) average eccentricity of
high (low) multiplicity samples: For the high multiplicity cut, the
KLN model (with the $p_T$ cutoff choices listed in Fig.~\ref{F5})
gives an average eccentricity $\langle
\epsilon_x\rangle{\,=\,}$0.03-0.04 (vs. $\langle
\epsilon_x\rangle{\,=\,}0.014$ from the Glauber model), whereas for
the low multiplicity cut the KLN model gives $\langle
\epsilon_x\rangle{\,=\,}$0.22-0.23 (vs. $\langle
\epsilon_x\rangle{\,=\,}0.24$ from the Glauber model).  We conclude
that the differences between the two classes of models are small, and
the suggested event selection scheme remains feasible in this new
framework. [It is worth noting that we have computed these same
eccentricity distributions for energies that will be available at the
LHC and found no significant deviations from the results presented
here.]

\begin{figure}[tb]
\includegraphics[width=0.45\textwidth,clip]{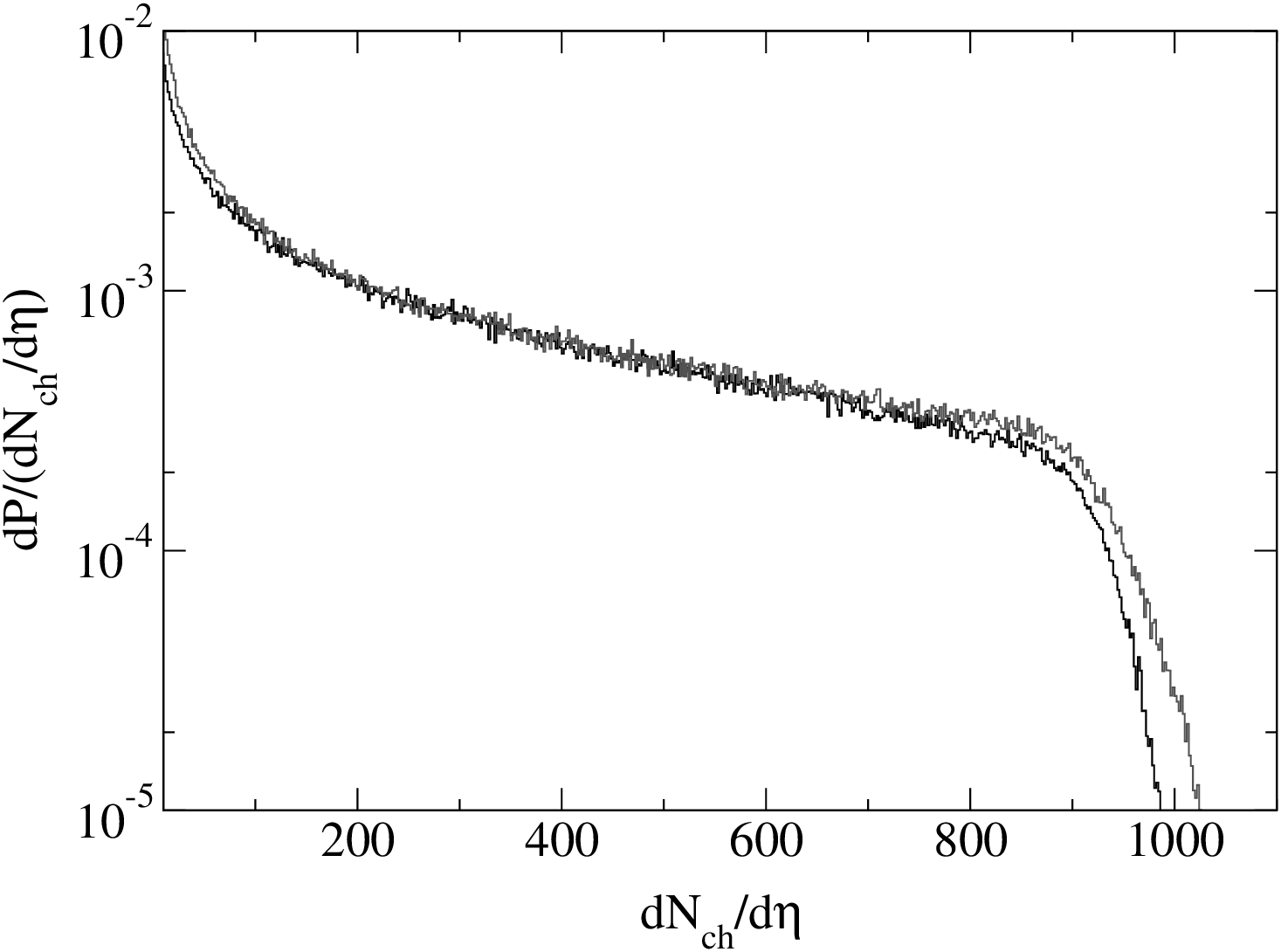}
\caption{Multiplicity distribution for U+U collisions with arbitrary
  impact parameters and relative orientations. The gray curve 
  corresponds to the Glauber model calculation of Ref.~\cite{Kuhlman:2005ts} 
  while the result from the KLN model is given by the solid curve.}
\label{F6}
\end{figure}

We now proceed to also discuss the effects of trigger inefficiencies
in the selection of full-overlap collisions on the eccentricity
selection. We repeat the calculations reported in
Ref.~\cite{Kuhlman:2005ts} for the KLN model, restricting our
attention to the case without $p_T$ cutoff
($\ptc{\,\to\,}\infty$). Figure \ref{F6} gives the resulting complete
multiplicity distribution for U+U collisions of any impact parameter
and relative orientation, and compares it with the previously studied
Glauber model initialization. The most notable feature is the reduced
width of the high-multiplicity tail in the KLN model, due to the
smaller spread of multiplicities in $b{\,=\,}0$, full-overlap
collision events seen in Fig.~\ref{F4}. The latter can be selected
from this distribution by placing very strict cuts on the number of
spectator nucleons. The gray histogram in Fig.~\ref{F4} shows the
multiplicity distribution for an event class that was obtained by
selecting from the KLN distribution in Fig.~\ref{F6} the 0.5\% events
with the lowest spectator counts. If we cut this multiplicity
distribution once more for high (low) multiplicities, we can again
select events with small (large) average eccentricity.

\begin{figure}
\includegraphics[width=0.45\textwidth,clip]{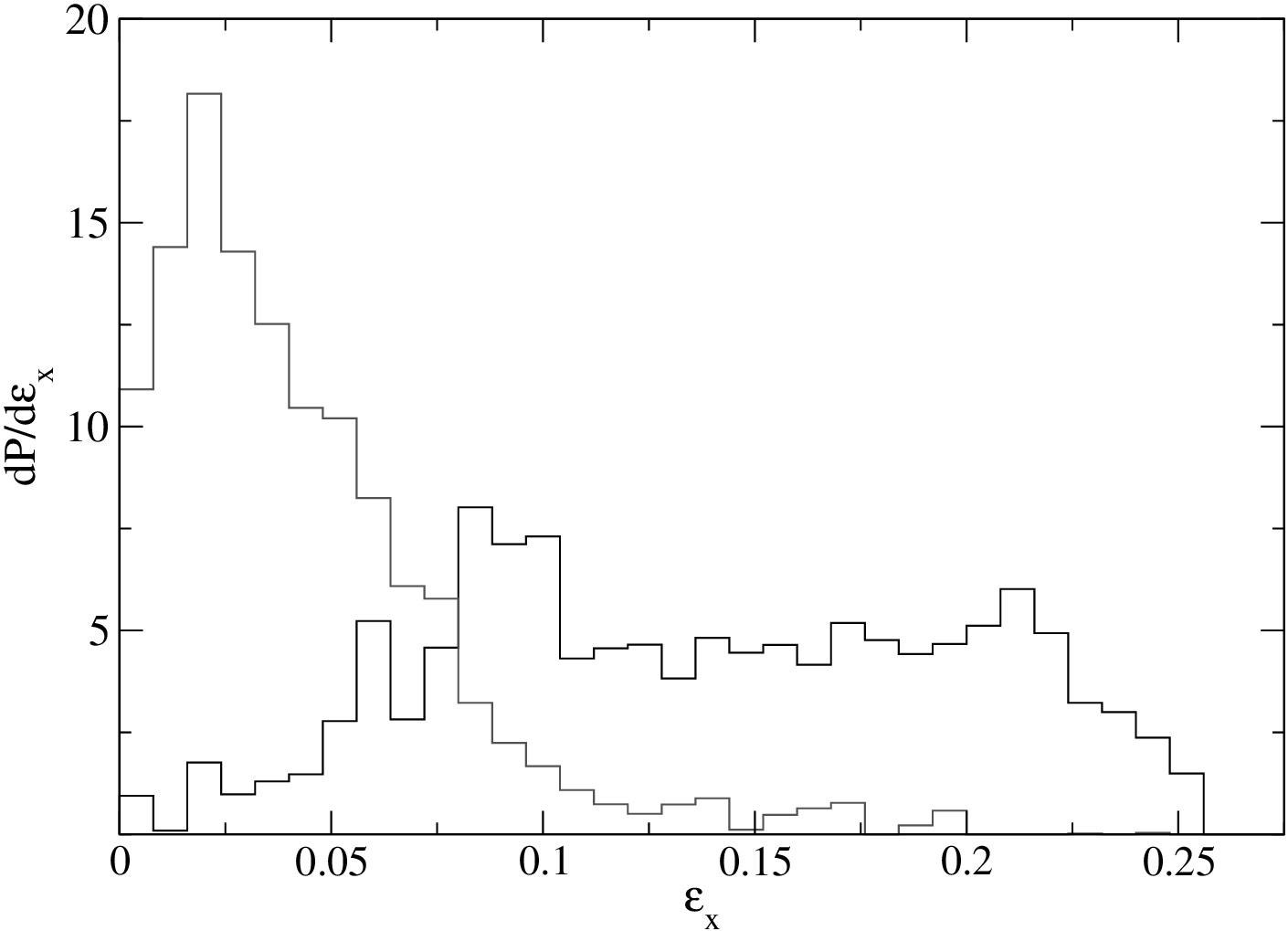}
\caption{Eccentricity distributions resulting from cutting the 
  gray histogram in Figure~\ref{F4} on the 5\% of events with the 
  lowest (black) and highest (gray) multiplicities. The corresponding 
  average eccentricities are 14\% (black) and 4\% (gray).}
\label{F7}
\end{figure}

The resulting eccentricity distributions are shown in Fig.~\ref{F7}.
The black (gray) histogram corresponds to the subclass of events 
obtained by selecting from the gray histogram in Fig.~\ref{F4} 
the 5\% lowest (highest) multiplicity events. While the eccentricities
are rather widely distributed within these two event subclasses,
their average eccentricities are very different: 
$\langle \epsilon_x\rangle{\,=\,}0.04$ for the high-multiplicity
subclass vs. $\langle \epsilon_x\rangle{\,=\,}0.14$ for the
low-multiplicity selection. (For the Glauber model the corresponding
average eccentricities were 0.02 and 0.18, respectively, and actual
values distributed somewhat more narrowly about these averages
\cite{Kuhlman:2005ts}.) 

We conclude that, in spite of the documented quantitative differences 
between the initial entropy distributions calculated from the KLN and 
Glauber models, the proposed event selection scheme for U+U collisions 
continues to allow isolating event classes with widely different source 
eccentricities, at energy and entropy densities that exceed those 
reachable in Au+Au collisions by a wide margin. Detailed comparisons
of elliptic flow measurements in U+U collisions with ideal fluid
dynamical calculations will, however, be affected by these model
uncertainties.

Before closing, we would like to point out the important role played 
by the collision geometry in determining from the saturation model the 
centrality dependence of particle multiplicity per participant. For 
simplicity, let us assume that the multiplicity of produced gluons at 
a given location $\bm{x}_\perp$ in transverse plane is given by
\begin{equation}
\label{dnde}
  \frac{dN}{d^2x_\perp \, dy} \, \propto \, 
  \min\{ Q_{s1}^2(\bm{x}_\perp), Q_{s2}^2(\bm{x}_\perp)\}.
\end{equation}
Here $Q_{s1}^2 (\bm{x}_\perp)$ and $Q_{s2}^2(\bm{x}_\perp)$ are the
saturation scales in the two colliding nuclei at the given transverse
position $\bm{x}_\perp$. Equation~\ref{dnde} captures with logarithmic
accuracy the main qualitative behavior of integrated particle 
multiplicity in the saturation/CGC approach. It can be obtained from 
\eq{kt_factorization} if one neglects logarithms of saturation scales 
as slowly-varying corrections to the power-law scaling of \eq{dnde}. 

\begin{figure}
\includegraphics[width=0.45\textwidth,clip]{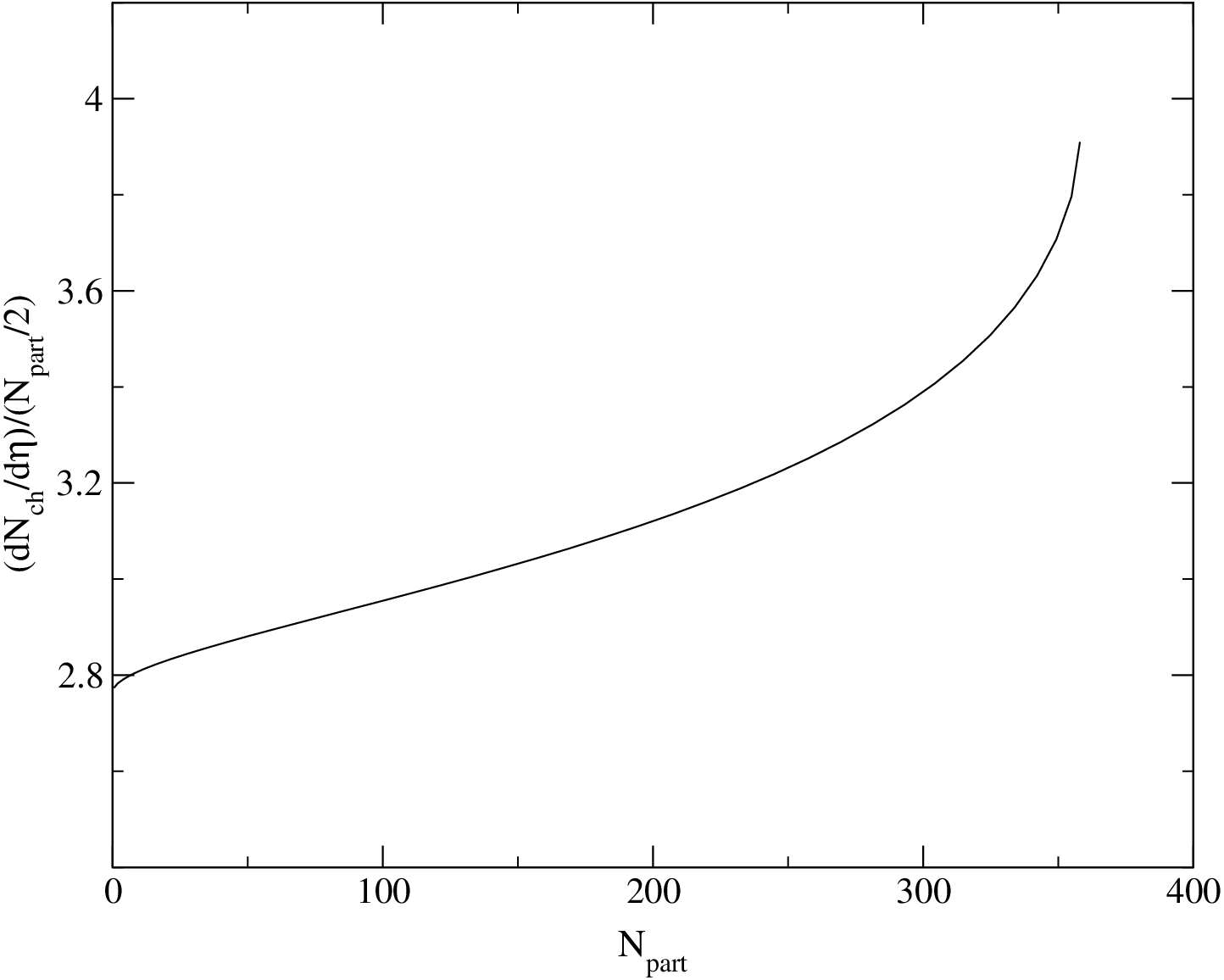}
\caption{Particle multiplicity per participant as a function of centrality,
for the simple CGC-motivated geometrical model \eq{dnde}.
The proportionality constant in \eq{dnde} is chosen to fit the most 
central data point from \cite{Back:2002gz} (see text for details).}
\label{F8}
\end{figure}

For illustration, let us consider a collision of two spherical nuclei 
(e.g. a Au+Au collision). If the radius of the nuclei is $R$ and they 
collide at impact parameter $\bm{b}$, the saturation scales are 
(assuming box-like nuclear density distributions for simplicity)
\begin{eqnarray}
\label{satsc}
 Q_{s1}^2 (\bm{x}_\perp) &\propto&
 \sqrt{R^2 - \bigl(\bm{x}_\perp{+}{\textstyle\frac{1}{2}}\bm{b})^2},
\nonumber\\
 Q_{s2}^2 (\bm{x}_\perp) &\propto&
 \sqrt{R^2 - \bigl(\bm{x}_\perp{-}{\textstyle\frac{1}{2}}\bm{b})^2},
\end{eqnarray}
with the same proportionality coefficient for both of them. Remembering 
that the square of the saturation scale is proportional to the density 
of participants $n_\mathrm{part}(\bm{x}_\perp)$, we deduce that the 
initial gluon (and thus also the final charged hadron) multiplicity 
per participant is proportional to
\be\label{mpp}
 \frac{1}{N_\mathrm{part}} \, \frac{dN_\mathrm{ch}}{dy} \, \propto \, 
 \frac{\int d^2x_\perp\,\min\{Q_{s1}^2(\bm{x}_\perp),Q_{s2}^2(\bm{x}_\perp)\}}
      {\int d^2x_\perp\,\bigl(Q_{s1}^2(\bm{x}_\perp)+Q_{s2}^2 (\bm{x}_\perp)
                        \bigr)},
\ee
with the total number of participant nucleons 
\be\label{Npart}
  N_\mathrm{part} \, \propto \, \int d^2x_\perp \, 
  \bigl(Q_{s1}^2(\bm{x}_\perp) + Q_{s2}^2(\bm{x}_\perp)\bigr).
\ee
The integrals in Eqs. (\ref{mpp}) and (\ref{Npart}) with the
saturation scales (\ref{satsc}) can be done numerically, yielding the
plot of particle multiplicity per participant as a function of
$N_{part}$ shown in \fig{F8}. The normalization of both axes was
adjusted to fit the most central data point in \fig{F2}. The curve in
\fig{F8} is seen to qualitatively reproduce the monotonous rise of
$(1/N_\mathrm{part})dN_\mathrm{ch}/dy$ with $N_\mathrm{part}$ observed
at RHIC \cite{Back:2002gz} (see \fig{F2}). We stress that the positive
slope of the curve in \fig{F8} arises entirely from the geometry of
the colliding nuclei, as implemented in the simple CGC-inspired model
\eq{dnde} for particle production. It is likely that a similar geometrical 
mechanism is responsible for the positive slope of the gluon
multiplicity per participant obtained in numerical simulations of
classical gluon fields performed in \cite{KNV}. Although often stated
otherwise, running coupling and/or other logarithmic effects appear
{\em not} to be necessary to obtain this positive slope, but they seem
to be required to turn the qualitative agreement of \fig{F8} with the
data in \fig{F2} into a quantitative one. Still, we are amazed how
well the data shown in \fig{F2} can be qualitatively understood within
the simple geometrical saturation model of gluon production given in
\eq{dnde}.

Although a number of different questions can be studied with U+U 
collisions \cite{Li00,Shuryak:1999by,KSH00,Heinz:2004ir}, perhaps the 
most persuasive argument in favor of a U+U program at RHIC is the 
potential \cite{Heinz:2004ir} to test the ideal hydrodynamic
behavior of the elliptic flow at higher entropy densities and with 
larger fireballs than are available in even the most central Au+Au 
collisions. In the Glauber model calculations presented in
\cite{Heinz:2004ir,Kuhlman:2005ts} we found that the maximum entropy 
density available in tip-on-tip U+U collisions was approximately 40\% 
greater than that seen in central Au+Au. Examination of Figure \ref{F1}
leads to a similar conclusion for the KLN model results, with increases 
of 35\% and 29\% for the cases with variable and infinite $p_T$ cutoff,
respectively. Since the initial maximum entropy density scales roughly
(although not exactly \cite{Nepali:2006va}) with the observable
$(1/\langle S\rangle)(dN_\mathrm{ch}/dy)$, where $\langle S\rangle$ is
the mean transverse overlap area of the two colliding nuclei, this 
increase in initial entropy density provides an important lever arm for
testing the approach towards ideal fluid dynamical behavior at RHIC 
experimentally. The additional parameter space opened up by U+U collisions 
at RHIC could prove even more important if Pb+Pb collisions at the Large 
Hadron Collider LHC (which are expected to produce {\em much} higher 
initial entropy densities and temperatures) turned out to produce a more 
weakly coupled quark-gluon plasma than RHIC collisions, and the QGP 
created at RHIC thus were to yield the best possible approximation of 
a ``perfect fluid''. In this case the U+U collision program could cover 
a key window in parameter space, but we also note that fully exploiting 
it would depend on our ability to gain sufficient control over the 
presently remaining model uncertainties for initial particle production, 
as studied here. 

\newpage

We wish to thank T.~Hirano and D.~Kharzeev for particularly illuminating 
discussions. This work was supported by the U.S.~Department of Energy 
under contract DE-FG02-01ER41190 (UH and AK) and OJI Grant 
No.~DE-FG02-05ER41377 (YuK).

\end{document}